\documentclass[reprint,
 amsmath,amssymb,
 aps,
prl,
]{revtex4-2}

\usepackage{graphicx}
\usepackage{dcolumn}
\usepackage{bm}
\usepackage{float}

\begin{document}

\preprint{APS/123-QED}

\title{Quantum logic control and precision measurements of molecular ions in a ring trap - a new approach for testing fundamental symmetries}

\author{Yan Zhou}
\email{yan.zhou@unlv.edu}
\affiliation{Department of Physics and Astronomy, University of Nevada, Las Vegas, Las Vegas, Nevada 89154, USA}
\author{Joshua O. Island}
\affiliation{Department of Physics and Astronomy, University of Nevada, Las Vegas, Las Vegas, Nevada 89154, USA}
\author{Matt Grau}
\affiliation{Department of Physics, Old Dominion University, Norfolk, VA 23529, USA}

\date{\today}

\begin{abstract}
We present a new platform facilitating quantum logic control of polar molecular ions in a segmented ring ion trap, paving the way for precision measurements. This approach focuses on achieving near-unity state preparation and detection, as well as long spin coherence. A distinctive aspect lies in separating state preparation and detection conducted in a static frame, from parity-selective spin-precession in a rotating frame. This method can be applied to a wide range of ion species and will be used to search for the electron’s electric dipole moment and the nuclear magnetic quadrupole moment.

\end{abstract}

\maketitle

The Standard Model of particle physics stands as a foundational framework for understanding the basic building blocks of matter and their interactions~\cite{THooft2007, Zyla2020}. Despite its success, the model leaves many questions unanswered, such as the nature of dark matter and dark energy, the mechanism behind the imbalance of matter and antimatter, and the lack of significant Charge and Parity (CP) violation in strong interactions~\cite{Engel2013, Feng2013a, Morrissey2012, Kim2010, Trodden1999, Kobayashi1973}. One approach to testing the Standard Model involves probing new particles generated in high-energy colliders~\cite{Wyatt2007}. Alternatively, more accessible, table-top experiments can be conducted using quantum sensors like polar molecules or highly charged ions to detect the dedicated interactions produced by new particles~\cite{Safronova2017, Ludlow2015, Kozlov2018}. This quantum sensor strategy is conceptually similar to electron-neutron scattering but operates on a much lower energy scale and within tightly bound atomic or molecular systems. In such systems, the probe electron in $s$-type orbitals spends a considerable amount of time in close proximity to and interacting with a heavy nucleus possessing relativistic energy, leading to a subtle but measurable frequency shift in electron spin resonance, which can be detected through precision spectroscopy~\cite{Andreev2018, Cairncross2019, Alarcon2022, Hutzler2020, Roussy2023}. 

In this letter, we propose a novel approach to measure the electron's Electric Dipole Moment (eEDM) by utilizing quantum logically controlled molecular ions in a segmented ring ion trap. The quantum logic scheme (QLS) allows for near-unity state preparation and detection~\cite{Chou2020, Micke2020, Chou2017, Wolf2016, Schmidt2005}. The circular motion of ions in the ring trap generates a rotating biased electric field of up to 32 V/cm and supports symmetry-violation searches through spin-precession measurements. To adapt the QLS method and spin-precession metrology, we develop a rotation-induced quantum control scheme that links the QLS procedure in a static frame with a DC magnetic field and the spin-precession in a rotating frame with synchronously rotating electric and magnetic fields. This scheme effectively decouples the quantum control and readout of molecular ions from the free evolution of spin-precession, thereby facilitating high efficiency in state preparation and detection as well as long coherence times simultaneously. Furthermore, this platform inherently supports temporally and spatially localized multiplexing measurements to suppress dominant systematic errors arising from inhomogeneous and time-variant magnetic fields. The proposed method incorporates many successful features from a well-established experimental platform developed by the JILA eEDM group~\cite{Meyer2009, Leanhardt2011, Loh2013, Cairncross2017, Gresh2016, Zhou2020,Ng2022, Roussy2023}. The newly designed experimental platform is anticipated to achieve a sub-$\mu$Hz accuracy equivalent to new physics at a mass scale above 100 TeV.

\begin{figure*}
\includegraphics[width=\textwidth]{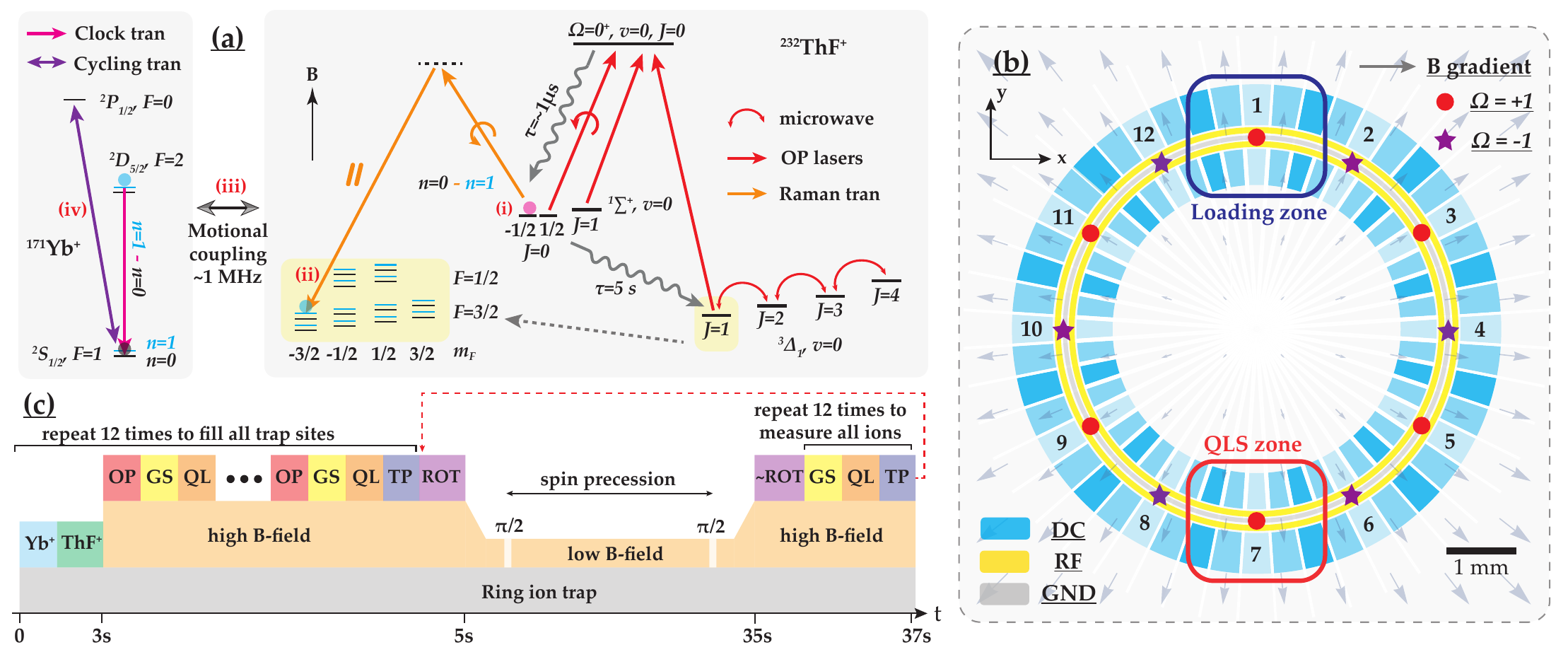}
\caption{\label{fig:exp} Experimental overview of the quantum logic control and spin-precession measurements in a segmented ring ion trap. (a) The schematic diagram of the state preparation of $^{232}$ThF$^+$ using optical-microwave pumping (OMP) and quantum logic spectroscopy (QLS) with Yb$^+$. (i) OMP concentrates around 30$\%$ population of $^{232}$ThF$^+$ in $^1\Sigma^+, v=0, J=0, m_\textrm{F}=-1/2$. Thereafter, (ii) QLS transfers this population to an eEDM sensitive state, denoted as $^3\Delta_1, v=0, J=1, m_\textrm{F}=-3/2$, $-$ parity. This is accomplished via a blue sideband of a two-photon transition, which excites one quanta in the collective motional mode of the $^{232}$ThF$^+$ and Yb$^+$ ion crystal (iii)~\cite{Schmidt2005,Chou2017}. The success of this transition, as evidenced by the motional excitation of the Yb$^+$ ion, is detected through an electron shelving process on the narrow $^2S_{1/2}$ to $^2D_{5/2}$ transition, and followed by resonant fluorescence on the $^2S_{1/2}$ to $^2P_{1/2}$ transition (iv). (b) A schematic diagram displays the segmented ring ion trap used for implementing the QLS scheme and precision metrology with biased electric and magnetic fields. The ions' circular motion generates a rotating electric field and synchronous magnetic field from a field gradient, which is generated by a pair of anti-Helmholtz coils (see Supplementary Materials FIG S3). (c) The time sequence of the precision measurements. The abbreviation terms in this plot include Yb$^+$, ThF$^+$: creating, trapping, and cooling Yb$^+$ and $^{232}$ThF$^+$. OP: optical pumping. GS: ground-state cooling. QL: quantum logic readout. TP: transport the ions from the state preparation site to an adjacent empty site. ROT: ramp up the rotation from 0 to 100 kHz. $\sim$ROT: ramp down the rotation from 100 kHz to 0. The dashed red arrow indicates the starting point for subsequent measurements.}
\end{figure*}

\textbf{State preparation and detection by QLS.}
The $^{232}$ThF$^+$ molecule, with its two valence electrons and closely spaced states of opposite parities, offers a significant eEDM enhancement in a weak bias electric field. However, the large number of nearly degenerate quantum states poses a challenge, as they dilute the desired coherent signal. The JILA eEDM group has demonstrated approximately 50$\%$ state preparation of $^{232}$ThF$^+$ in an eEDM-sensitive state ($^3\Delta_1$, $v$=0, $J$=1, and single $m_\textrm{F}$) by employing optical and microwave pumping (OMP)~\cite{Zhou2020}. Moreover, they have developed a rotation-selective resonant enhanced multiphoton dissociation (REMPD) detection scheme with an efficiency around 30$\%$~\cite{Ni2014}. The impurity in state preparation stems from population leakage to low-lying electronic states and vibrationally excited states ($v>1$). The non-ideal efficiency of the dissociation detection could be attributed to the complexity of multiple dissociative and radiative channels of the highly excited $^{232}$ThF$^+$. To improve the state preparation and detection, substantial spectroscopic work and additional repumping lasers are required. Furthermore, these efforts necessitate repetition and may exhibit significant variation for different molecular species. In this letter, we propose a universal method that combines OMP and QLS for single quantum state preparation and detection with near-unity efficiency. This approach can be directly applied to more complex molecular ions, such as $^{229}$ThF$^+$, $^{181}$TaO$^+$, and $^{176}$LuOH$^+$ without substantial strategy modifications. (see Supplementary Materials Section 1 for more details). 

FIG \ref{fig:exp}(a) illustrates our state-preparation procedure. \textbf{First}, $^{232}$ThF$^+$ ions are generated by resonance-enhanced multi-photon ionization (REMPI), by which the population is distributed in a single vibronic state but four rotational states. \textbf{Second}, one $^{232}$ThF$^+$ and one Yb$^+$ are loaded into a radiofrequency (RF) ion trap at loading zone (trap site 1 in FIG \ref{fig:exp}(b), see Supplementary Materials Section 2 for more details) and Doppler cooled. \textbf{Third}, OMP transfers approximately 30$\%$ of $^{232}$ThF$^+$ ions to $^1\Sigma^+$, $v=0$, $J=0$, $m_\textrm{F}=-1/2$ state. Spin polarization is generated by a static magnetic field and a circularly polarized pumping laser. \textbf{Fourth}, both $^{232}$ThF$^+$ and Yb$^+$ ions are transferred to the QLS zone (trap site 7 in FIG \ref{fig:exp}(b)) by modulating the axial trapping potential. Ground state cooling and electron shelving are applied to Yb$^+$, preparing it in $^2D_{5/2}$, $F=2$, $n=0$ state, where $n$ labels the motional state quantum number of the Yb$^+$ and $^{232}$ThF$^+$ ion crystal in the harmonic trap. \textbf{Fifth}, a pair of referenced Raman pulses coherently transfer $^{232}$ThF$^+$ to one eEDM sensitive state, such as $^3\Delta_1$, $v=0$, $J=1$, $m_\textrm{F}=-3/2$, $-$ parity, with motional excitation ($n=1$). The $m_\textrm{F}$ selectivity is due to the polarization of the Raman beams and Zeeman detuning. After OMP in the third step, if $^{232}$ThF$^+$ is not in the desired state ($^1\Sigma^+$, $v=0$, $J=0$, $m_\textrm{F}=-1/2$), the motional excitation through the Raman transfer process will not happen, and the normal mode of the ion crystal will still be $n=0$. \textbf{Sixth}, a blue sideband $\pi$ pulse ($n=1 \rightarrow n=0$) resonantly de-excites the Yb$^+$ ion's electronic and motional states to $^2S_{1/2}$, $F=1$, $n=0$, allowing for cycling fluorescence detection. If the detected photons exceed a threshold, according to the motional entanglement between Yb$^+$ and $^{232}$ThF$^+$, it is nearly certain that the $^{232}$ThF$^+$ is at the eEDM sensitive state. Otherwise, the above process is repeated until the conditions for state preparation are met. 

Drawing from the greater than 30$\%$ OMP efficiency demonstrated by the JILA eEDM group without any vibration repumps~\cite{Zhou2020}, it's feasible to attain unity eEDM-sensitive state preparation of $^{232}$ThF$^+$ within just a few iterations. Even for more complex molecules with larger nuclear spins, such as $^{229}$ThF$^+$, although the OMP method alone may result in lower state preparation efficiency, implementation of QLS can offset this through additional iterations. State detection involves a Raman transfer of $^{232}$ThF$^+$ from the $^3\Delta_1$ to $^1\Sigma^+$ state, employing a similar QLS readout method.

\textbf{Segmented ring ion trap and rotating electric and magnetic fields.}
Probing fundamental symmetry breaks using molecular ions requires the application of a biased electric field to mix states with opposite parities. However, trapped ions are typically incompatible with such electric fields, as even a weak DC electric field (around 10 V/cm) can eject the ions from the trap. The JILA eEDM group first demonstrated polarization of molecular ions using a rapidly rotating electric field~\cite{Loh2013, Cairncross2017, Roussy2023}. Implementing QLS in a rotating frame, however, presents challenges due to light delivery complications and excess heating. In this letter, we resolve this conflict by entirely decoupling the QLS and spin-precession. A key of our solution involves designing and fabricating a segmented ring trap, as depicted in FIG \ref{fig:exp}(b). This trap incorporates several features from a design created by Sandia National Laboratories, which is intended for scalable quantum computation~\cite{Tabakov2015}. 

The ring trap has a radius of 3 mm, with ions situated approximately 50 $\mu$m above the surface. The tight radial confinements ($\sim$2.5 MHz) are achieved through three concentric RF (20 MHz, 200 V$_\textrm{pp}$) and ground electrodes, while the loose axial confinement ($\sim$1 MHz) comes from the segmented DC electrodes inside and outside the RF ring electrodes. The trap consists of 96 segmented electrodes divided into 12 groups (+V, 0, -V, 0), forming 12 trapping sites. The trapping sites and ions inside can be rotated clockwise or counterclockwise by modulating the segmented electrodes sinusoidally. Centripetal acceleration provides a rotating electric field that polarizes polar molecules. The maximum amplitude of the electric field (32 V/cm) is determined by the trap radius and maximum rotating frequency (100 kHz). A radial magnetic field gradient is applied by a pair of anti-Helmholtz coils concentric to the ring trap (see Supplemental Materials, Fig. S3). As the ions rotate, they also experience a rotating magnetic field that is either parallel or anti-parallel to the electric field. Reversing the coil current changes the magnetic field direction. When the ions are stationary, although the net electric field is zero, a non-zero static inward/outward magnetic field remains, defining the quantization axis for the QLS scheme of state preparation and detection. 

The fabrication of the proposed trap can be achieved using standard microfabrication techniques. Key aspects of the fabrication process include creating a planarized surface for the trap electrodes and avoiding obstructions that may protrude from the trap surface. To obtain a flat surface devoid of bonding wires that could interfere with lasers and trapped ions, silicon plug/via technology will be utilized~\cite{Tabakov2015, Mielenz2016, Zhao2021, Li2017}. Employing CMOS technology in this manner enables the fabrication of multilayer structures with both RF and DC electrical access~\cite{Blain2021}.

\begin{figure}[b]
\includegraphics[width=0.48\textwidth]{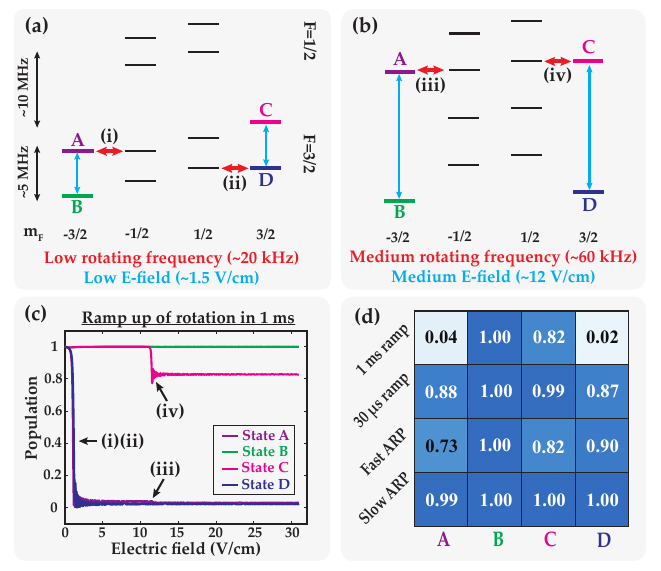}
\caption{\label{fig:spin} Spin-polarization during the rotating ramp-up procedure. Panels (a) and (b) show degenerate rotational couplings at low and intermediate electric fields. The light blue arrows mark Stark interactions, and the red arrows mark rotation couplings. Panel (c) shows numerical evaluations of spin-polarization dynamics with a 1 ms linear ramp. The arrows mark the population leakage induced by rotational couplings that are indicated in Panels (a) and (b). Panel (d) compares different rotating ramp-up schemes.}
\end{figure}

\textbf{Spin-polarization during rotation ramped-up.}
By integrating QLS, segmented ring trap, and static quadrupole magnetic field, we illustrate the state preparation of $^{232}$ThF$^+$ in a single spin-polarized eEDM sensitive state (A, B, C, or D in FIG \ref{fig:spin}). Thereafter, the stationary ions will be transferred to a fast-rotating frame, experiencing a net electric field. However, during this process, the rotation-induced coupling, $H_\textrm{rot}$, may mix neighbouring Zeeman sub-levels ($\Delta m_\textrm{F}$ = $\pm$1) and consequently depopulate the target states~\cite{Meyer2009}:
\begin{equation}
 H_\textrm{rot}=-\omega_\textrm{rot}(F_z\cos{\theta}-F_x\sin{\theta}),
\end{equation}
where $\theta$ is the angle between the rotation and quantization axes and equal to 90$^\circ$ in this case, $\omega_\textrm{rot}$ is the rotational angular frequency. The interaction can be strong when the perturbed states ($m_\textrm{F}$=$\pm$1/2) become degenerate to the spin-polarized states ($m_\textrm{F}$=$\pm$3/2), due to cancellations of Zeeman and Stark shifts at around 1.5 V/cm and cancellations of Stark and hyperfine shifts at around 12 V/cm electric fields respectively, as indicated by the red arrows in FIG \ref{fig:spin}(a) and (b). We evaluate the population transfer during the rotating ramp-up induced by such degenerate interactions. FIG \ref{fig:spin}(c) reveals that the rotational coupling entirely depopulates states A and D, and partially depopulates state C with a 1 ms linear ramp. Only the population of state B remains fully preserved. 

To suppress the spin-depolarization caused by degenerate interactions, we can increase the rotating ramp rate. Our model shows that a 30 $\mu$s ramp duration (3.3 MHz/ms ramp rate) can preserve the spin-polarization by approximately 90$\%$. However, applying such a rapidly accelerating field may either introduce detrimental axial electric field or excessive heating or even ion detrapping. A more feasible approach starts with the population in non-spin-polarizing states, and then adiabatically transfers them to the target spin-polarized states through slowly ramped rotation frequency ($<$2 kHz/ms ramp rate). FIG \ref{fig:spin}(d) summarizes the spin-polarization efficiency after rotation ramp-up using different methods. Through the slow adiabatic rapid passage (ARP) within 75 ms, nearly 100$\%$ populations are prepared in the target states. Calculation details are described in the Supplementary Materials Sections 4 and 5. 

\textbf{Coherent superpositions.}
To initiate and terminate the spin-precession, it is crucial to generate effective $\pi/2$-pulses between states A and C, or B and D. Following the approach initiated by the JILA eEDM group, we can establish such coherence by leveraging a fourth-order interaction resulting from combined perturbations of rotational coupling and Stark mixing~\cite{Cairncross2017}. An effective two-level Hamiltonian of states A and C, or B and D, can be formulated as:
\begin{equation}
\resizebox{0.95\hsize}{!}{
$H = \left[\begin{matrix}-3g_\textrm{F}\mu_\textrm{B}B&p\left(\frac{\omega_\textrm{rot}}{E_\textrm{rot}}\right)^3\\p\left(\frac{\omega_\textrm{rot}}{E_\textrm{rot}}\right)^3&+3g_F\mu_\textrm{B}B\\\end{matrix}\right]
=\left[\begin{matrix}-3g_F\mu_\textrm{B}B&p^{\prime}\left(\frac{1}{\omega_\textrm{rot}}\right)^3\\p^{\prime}\left(\frac{1}{\omega_\textrm{rot}}\right)^3&+3g_\textrm{F}\mu_\textrm{B}B\\\end{matrix}\right]$
\nonumber}
\end{equation}

where $g_F$ is the $g$-factor of the $F=3/2$ state, $\mu_\textrm{B}$ is the Bohr magneton, $B$ is the amplitude of the effective rotating magnetic field (${\bf{r}} \cdot \nabla {\bf{B}}$), and all other parameters are grouped in $p$ and $p^{\prime}$. The rotating electric field is determined by $E_\textrm{rot}=m\omega_\textrm{rot}^2r/e$, where $r$ corresponds to the trap radius and $e$ represents a unit charge. Consequently, the diagonal matrix elements can be manipulated by altering the applied magnetic field gradient, while the off-diagonal matrix elements can be regulated by adjusting the rotating frequency. This enables the generation of off-resonant $\pi/2$ pulses, either by modulating the magnetic field gradient or the rotating frequency. The former method is favored due to its preservation of spin-polarization.

\textbf{Experimental sequence.}
Incorporating all the above-described procedures, FIG \ref{fig:exp}(c) illustrates a comprehensive time sequence for the measurements. \textbf{First}, $^{232}$ThF$^+$ and Yb$^+$ ions are generated and trapped. \textbf{Second}, the QLS state preparation scheme is employed to produce a single state of $^{232}$ThF$^+$ in a static magnetic field. To ensure the unity state preparation, multi cycles of optical pumping (OP), ground state cooling (GS), and quantum logic readout (QL) may be required. \textbf{Third}, the $^{232}$ThF$^+$ and Yb$^+$ are transported to other trap nodes (TP). It is anticipated that twelve $^{232}$ThF$^+$ ions will be prepared within two seconds. These ions may be prepared in the same quantum state or interleaved in different states to suppress systematic errors. \textbf{Fourth}, a slow rotation ramp transfers all ions from a static frame to a 100 kHz rotating frame with a 32 V/cm bias electric field. \textbf{Fifth}, the magnetic field amplitude is decreased from 10 gauss to 10 milli-gauss, followed by abruptly turning the magnetic field off and on for approximately 100 ms to induce coherence (effective $\pi$/2 pulse) between the spin-polarized states. \textbf{Sixth}, phase accumulation occurs during an extended spin-precession period in the rotating frame. Subsequently, another $\pi$/2 pulse is applied to map the phase information to the population difference. \textbf{Seventh}, steps 2 to 5 are reversed – the magnetic field is ramped up, the rotation frequency is reduced, and the QLS method is applied to read out states of all twelve ions. \textbf{Finally}, assuming the absence of incoherent interactions to the environment after quantum state readout, $^{232}$ThF$^+$ should keep in pure spin-polarization states. As a result, subsequent spin-precession measurements can commence immediately without the necessities of state preparation. 

\textbf{Systematics.}
Following the JILA eEDM experiments, three pivotal binary switches will be employed: the direction of magnetic field ($B$ switch), the direction of rotation ($R$ switch), and the molecular orientation ($D$ switch)~\cite{Cairncross2017, Roussy2023}. To eliminate prominent systematic terms, a linear combination of eight independent measurements will be utilized. In our scheme, twelve ions in the ring trap will be prepared with different orientations, interlaced with the $D$ switch. Spin precession measurements featuring opposite effective electric fields will be conducted simultaneously in both time and space. Such spatially and temporally parallel measurements serve to suppress systematics arising from drifting and inhomogeneous electric and magnetic fields, particularly when the coherence time experiences a significant increase.

\textbf{Scalability and sensitivity.}
In contrast to precision measurements employing ensembles of molecules, utilizing a single molecule per trap site compromises the signal readout. To counteract the reduced signal count, it is crucial to develop a scalable ring trap capable of accommodating numerous trap sites. Additionally, the integrated photonic technology pioneered by MIT Lincoln Laboratory may be leveraged to scale light delivery to the trap~\cite{Niffenegger2020}. As a result, the practical implementation of a large ring trap comprising hundreds of trap sites becomes feasible. Moreover, the absence of high-power pulsed lasers, unstable molecular sources, or mechanically moving parts facilitates long-term data acquisition through the incorporation of an autopilot control system.

Table \ref{tab:results} presents the estimated statistical sensitivities of this scheme. The frequency uncertainty is determined by:
\begin{equation}
d\nu = \frac{1}{2 \pi C \sqrt{N \tau T D P R}}
\end{equation}
The parameters in the above equation are described in Table \ref{tab:results}. Generation I experiments, conducted at room temperature, could aim to achieve three primary objectives: (1) demonstrate near-unity state preparation and readout schemes utilizing QLS, (2) validate the proposed precision metrology, and (3) conduct an initial investigation of systematic errors. Generation II could be performed at cryogenic temperature to maximize the coherence time through increasing ion trap storage times by improved vacuum in the cryogenic environment, and suppressing rotational and vibrational state changing excitation from black-body radiation. Generation III would focus on implementing highly multiplexed measurements and long data acquisition duration for an ultimate statistical sensitivity. Compared to the leading eEDM measurement by the JILA group using HfF$^+$ with 22.8 $\mu$Hz sensitivity~\cite{Roussy2023}, Generation I would reach a similar level, and Generation III would improve the current best measurement by about two orders of magnitude. 

In this letter, we present a design of a novel experimental platform for CP-violation measurements utilizing quantum logically controlled molecular ions. While the primary focus is on eEDM measurements using $^{232}$ThF$^+$, this scheme will be applied to probe nuclear magnetic quadrupole momentum using $^{229}$ThF$^+$, $^{181}$TaO$^+$, and $^{176}$LuOH$^+$.

\begin{table}
\begin{center}
\begin{tabular}{c|c|c|c} 
\hline\hline
 Parameters & Gen I & Gen II & Gen III \\
\hline
Temperature & 300 K & 4 K & 4 K \\
No. of trap sites ($N$) & 12 & 12 & 400 \\ 
State preparation ($P$) & 90$\%$ & 95$\%$ & 95$\%$ \\ 
State readout ($R$) & 90$\%$ & 95$\%$ & 95$\%$ \\
Contrast ($C$) & 90$\%$ & 95$\%$ & 95$\%$ \\
Coherence time ($\tau$) & 4 s & 100 s & 100 s \\
Duty cycle ($D$) & 50$\%$ & 95$\%$ & 90$\%$ \\
Data collection ($T$) & 300 h & 300 h & 1000 h\\
\hline
Frequency precision & 39 $\mu$Hz & 5 $\mu$Hz & 0.5 $\mu$Hz \\
\hline\hline
\end{tabular}
\end{center}
\caption{\label{tab:results} Target sensitivities of Generations I, II, and III.}
\end{table}

\begin{acknowledgments}
We thank T. N. Taylor, B. Zygelman, A. Jayich, C. W. Chou, A. Bradley, and R. W. Field for helpful discussions. This work is financially supported by UNLV Faculty Opportunity Award and by the Virginia Space Grant Consortium.  
\end{acknowledgments}

\bibliography{apssamp}

\clearpage
\onecolumngrid
\appendix
\begin{center}
\large \textbf{Supplemental materials for ``Quantum logic control and precision measurements of molecular ions in a ring trap - a new approach for testing fundamental symmetries"}
\end{center}
\hfill

\begin{center}
Yan Zhou$^{*1}$, Joshua O. Island$^1$, Matt Grau$^2$

\textit{$^1$Department of Physics and Astronomy, University of Nevada, Las Vegas, Las Vegas, Nevada 89154, USA}

\textit{$^2$Department of Physics, Old Dominion University, Norfolk, VA 23529, USA}
\end{center}

\renewcommand\thefigure{S\arabic{figure}}
\renewcommand\thetable{S\arabic{table}}
\setcounter{figure}{0} 
\setcounter{page}{1}

\section{Section 1: State preparation without optical pumping}
In this letter, we describe a method that integrates optical and microwave pumping (OMP) and quantum logic scheme (QLS) to prepare $^{232}$ThF$^+$ into a single quantum state. The flexibility of the QLS method allows it to be directly extended to a broad spectrum of molecular ions, including but not limited to $^{229}$ThF$^+$, $^{181}$TaO$^+$, and $^{176}$LuOH$^+$. These molecular ions, in particular, exhibit significant enhanced sensitivity, thereby enabling more investigations into the fundamental symmetries within the hadronic sector of the Standard Model. However, the OMP procedure hinges largely on exhaustive spectroscopic understanding of these molecules, a criterion that remains unfulfilled at present. Compounding this issue are the intricate molecular structures stemming from the substantial nuclear spins of the heavy elements. This complexity inevitably undermines the OMP efficiency, particularly given its reliance on nearly-closed-cycle optical transitions, resulting in an efficiency decrease of one to two orders of magnitude~[S1]. To overcome this issue, we have designed an adaptation of the state preparatio  n process, making it independent on the stringent requirements of the OMP technique. This adaptation enables its applicability to a wider range of molecular ions. 

\begin{figure}[h]
\includegraphics[width=0.4\textwidth]{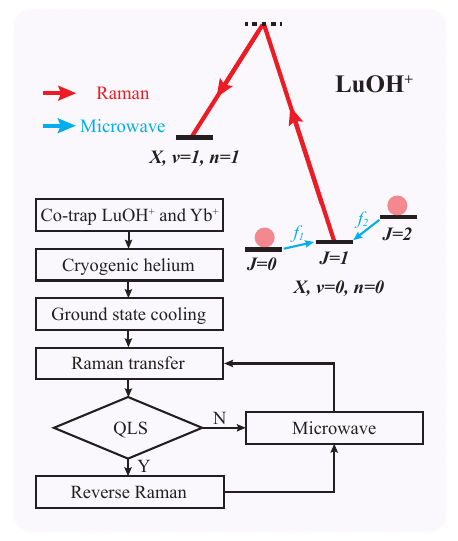}
\caption{\label{fig:QLMS} Quantum Logic Microwave Scheme (QLMS) for single state preparation of $^{176}$LuOH$^+$. This scheme operates independently of the spectroscopic details of the ions' excited electronic states.}
\end{figure}

FIG \ref{fig:QLMS} illustrates the adaptive state-preparation protocol. The process starts with the employment of cryogenic helium gas, which is utilized to sympathetically concentrate molecular ions from thousands of initial states at room temperature to a few hundred states at a 4K temperature. Subsequently, microwave pulses couple every potentially populated state to the desired state in a sequential manner, each with its distinct resonant frequency. Such information can be obtained precisely by spectroscopy of the molecular ground electronic state only. Following each iteration of microwave coupling, the QLS detection is applied to determine whether the target molecular ion is at the desired state. If the QLS procedure generates a positive signal, indicating that the molecular ion is in the desired state, the state preparation procedure is complete. Otherwise, additional microwave pulses are deployed to transfer the population from other states to the desired state until a positive QLS detection is observed. Under the most demanding circumstances, with the the population initially distributed across several hundred states after cryogenic cooling, the aforementioned procedure might need to be repeated hundreds of times, taking up to several tenth of seconds. The development and implementation of an optimized and dynamic microwave pulse sequence, informed by previous negative QLS measurements, could significantly reduce the number of trial iterations and speed up the state preparation process~[S2]. 

\section{Section 2: Molecular creation and loading into the ring trap}
\begin{figure}[h]
\includegraphics[width=0.6\textwidth]{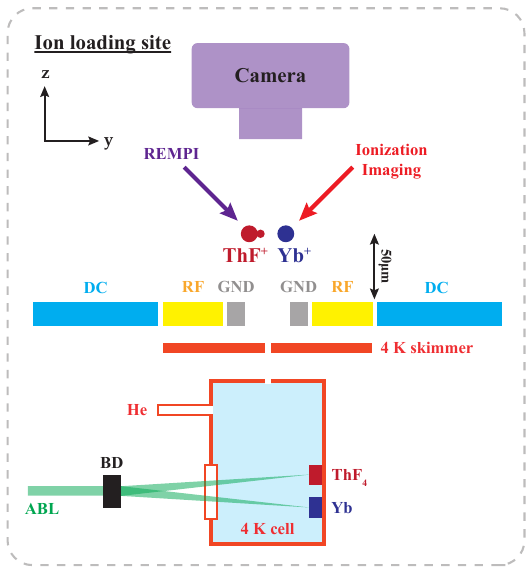}
\caption{\label{fig:load} Schematic diagram of the ion loading zone from side view. The abbreviation terms in this plot include BD: beam deflector, ABL: ablation laser, REMPI: resonance enhanced multiphoton ionization. }
\end{figure}

To alleviate the congestion of laser beams in the system, we propose separating the ion loading zone from the QLS zone. FIG \ref{fig:load} presents a detailed side view (in the YZ-plane) of the ion loading area, also referred to as trap site 1. In this arrangement, a slot measuring 20 $\mu$m in width is carefully incised into the central ground ring. Beneath trap site 1, we set up a cryogenic buffer gas cell, which is filled with 4K helium gas. The cold cell is equipped with pellets of Thorium tetrafluoride ($^{232}$ThF$_4$) and Ytterbium metal. Upon heating by a pulsed laser directed by a motorized beam deflector, neutral $^{232}$ThF molecules and Yb atoms are vaporized and subsequently cooled via collisions to the cryogenic helium gas. After that, they exit the cell through an aperture at the top. To prevent contamination of the uncollimated beam originating from the buffer gas cell, a cold skimmer with a diameter of approximately 20 micrometers is placed between the ring trap and the cold cell.

When the Yb atoms reach the trap site, which is approximately 50 micrometers above the electrode plane, they undergo photoionization and are subsequently imaged by lasers. If a single Yb$^+$ ion is successfully loaded, the Yb ionization laser is immediately switched off. Following this, the resonantly enhanced multi-photon ionization (REMPI) lasers, which are responsible for ionizing $^{232}$ThF, are activated for the purpose of loading a $^{232}$ThF$^+$ ion. To oversee the process, a CCD camera is positioned at the top of the trap. This monitors the Yb$^+$ ion's position by capturing fluorescence imaging. The successful loading of a single $^{232}$ThF$^+$ ion will cause the Yb$^+$ ion to shift by a precisely determined distance. 

\section{Section 3: Rotating electric and magnetic fields}
\begin{figure}[h]
\includegraphics[width=0.9\textwidth]{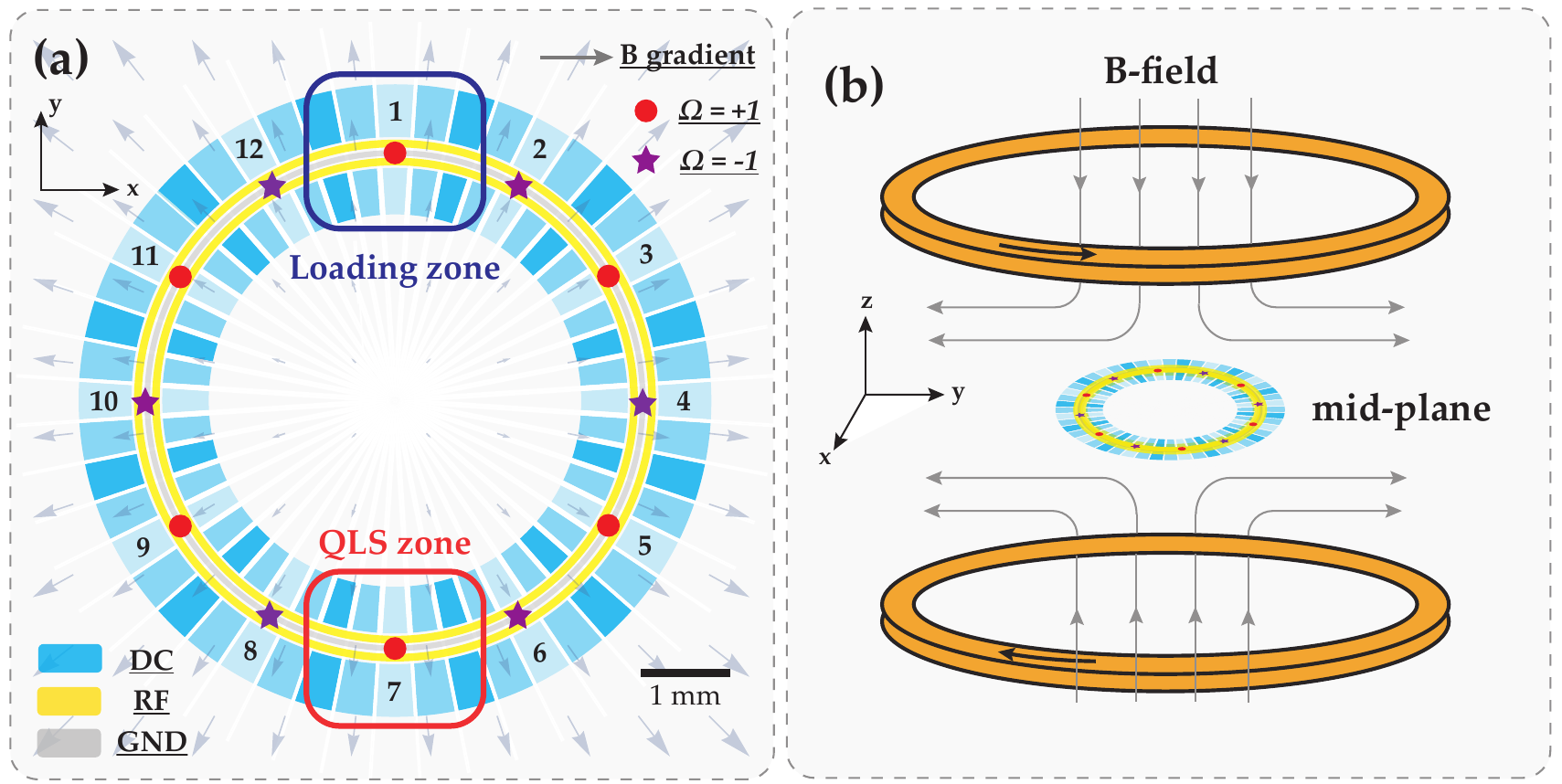}
\caption{\label{fig:coils} (a) Schematic diagram of the segmented ring ion trap for implementing the QLS scheme in the precision measurements with electric bias field. The trap is composed of axial control electrodes which have the capability to maneuver the ions within the trap in either a clockwise or counter-clockwise direction. (b) A pair of anti-Helmholtz coils is implemented to generate a linear radial magnetic gradient within the plane of the trap. This configuration ensures the generation of a magnetic field that is synchronized to the electric field in the rotating frame. By reversing the driving current, we can reverse the direction of the magnetic field.}
\end{figure}

In our attempt to synchronize the generation of a rotating magnetic and electric fields as ions circulate within the ring trap, we employ a strategy inspired by the JILA eEDM group~[S3]. This method leverages a synergistic combination of a static magnetic field gradient and the naturally occurring rotational motion of ions. In the context of our experimental design, we introduce a radial magnetic field gradient, expressed as ${\bf{B}} = {B'_\textrm{rad}}\left( {2{\bf{Z}} - {\bf{X}} - {\bf{Y}}} \right)$, which is produced by a pair of anti-Helmholtz coils. The coils are arranged concentrically respect to the ring trap in the XY-plane and are symmetric about the Z-axis, as shown in FIG \ref{fig:coils}. The magnitude of the radial magnetic field gradient, $B'_\textrm{rad}$, typically ranges from 0.01 to 10 G/cm. Consequently, the ions rotate with a fixed offset from the trap center, traversing a magnetic field of constant magnitude but a rapidly rotating direction, which is either parallel or anti-parallel to the electric field in the rotating frame. The amplitude of the rotating magnetic field is determined by ${B_\textrm{rot}} = {B'_\textrm{rad}}{r_\textrm{rot}}$. In the context of this experimental design, ${B_\textrm{rot}}$ typically ranges from 3 mG to 3 G, which corresponds to Zeeman shifts from 50 Hz to 50 kHz. By reversing the coil current, we can flip the magnetic field direction relative to the electric field, which is critical for eEDM extraction and systematic error analysis. A noteworthy deviation from the JILA eEDM experiment is that the $r_\textrm{rot}$ remains constant and is independent of ion motion. Hence, even with stationary ions within the ring trap, a non-zero static inward/outward magnetic field persists. This static magnetic field plays a crucial role as it establishes the quantization axis for the QLS scheme, which is responsible for the state preparation and detection. 

\section{Section 4: Effective Hamiltonian of $^{232}$T\lowercase{h}F$^+$ $^3\Delta_1$ states}

The calculations for spin-polarization during the rotation ramp-up process are based on an effective Hamiltonian of $^3\Delta_1$ state of $^{232}$ThF$^+$~[S3,S4]. The $^3\Delta_1$ state is typically described using Hund's case (a) basis states with coupled nuclear spin,
\begin{equation}
\left| {\Lambda  =  \pm 2,S = 1,\Sigma  =  \mp 1,J,\Omega  =  \pm 1,I = 1/2,F,{m_\textrm{F}}} \right\rangle, 
\label{eqn:basis}
\end{equation}
\noindent where $\Lambda  = {\bf{L}} \cdot {\bf{n}}$ is the projection of the electronic orbital angular momentum ($\bf{L}$), onto the molecular inter-nuclear axis ($\bf{n}$). The electronic spin angular momentum is represented by $S=|\bf{S}|$. The projection of this electronic spin on the molecular inter-nuclear axis is $\Sigma = \bf{S} \cdot \bf{n}$. $J=|\bf{J}|=|\bf{L}+\bf{S}+\bf{R}|$ is the angular momentum, which accounts for electron ($\bf{L}+\bf{S}$) and molecular rotation ($\bf{R}$). $\Omega=\bf{J} \cdot \bf{n}$ is the projection of the total electronic angular momentum ($\bf{L}+\bf{S}$) onto the molecular inter-nuclear axis. $\bf{R}$ does not contribute to $\Omega$ because $\bf{R} \cdot \bf{n} = 0$. The nuclear spin of $^{19}$F is denoted by $I=|\bf{I}|$, while the nuclear spin of $^{232}$Th is zero. $F=|\bf{F}| = |\bf{J}+\bf{I}|$ stands for the total angular momentum, and $m_\textrm{F}=\bf{F} \cdot \bf{z}$ represents the projection of this total angular momentum on the rotating quantization axis, which is defined by the electric field. We define the direction of the molecular inter-nuclear axis $\bf{n}$ as extending from the $^{19}$F nucleus towards the $^{232}$Th nucleus. 

In this letter, our primary objective is to create a model that accurately captures the dynamics of spin-polarization during the rotation ramp-up process. This task is substantially less intricate than investigating eEDM systematic errors. Thus, we streamline our effective Hamiltonian to encapsulate only key interactions of relevance, which compromises: the nuclear spin hyperfine interaction, the Stark effect, $\Omega$-doubling, rotation induced $m_\textrm{F}$ coupling, and the Zeeman effect: 
\begin{equation}
H\left( {{\bf{E}},{\bf{B}},{\omega _\textrm{rot}}} \right) = {H_\textrm{hf}} + {H_\textrm{S}}({\bf{E}}) + {H_\Omega}  + {H_\textrm{rot}}({\omega _\textrm{rot}}) + {H_\textrm{Z,e}}({\bf{B}}) + {H_\textrm{Z,N}}({\bf{B}}),
\label{eqn:Hamiltonian}
\end{equation}
\noindent where $\bf{E}$ and $\bf{B}$ denote vector representations of the electric and magnetic fields in the rotating frame, respectively. $\omega_\textrm{rot}$ is the angular frequency at which the the molecular ions circulate within the ring trap. The arrangement of the terms in Eq. \ref{eqn:Hamiltonian} is based on energy scales prevalent under typical experimental conditions. Quantum states that are substantially far off-resonance, such as those relating to excited electronic, vibrational, and rotational states, are omitted from our consideration. Furthermore, we disregard interactions that have minimal impact on the spin-polarization process, such as eEDM. 

For each term in Eqn. \ref{eqn:Hamiltonian}, we first express them in terms of effective operators, and then evaluate the matrix elements of these operators using the Hund's case (a) basis. The nuclear magnetic hyperfine structure from $I=1/2$ nuclear spin of the Fluorine nucleus is:
\begin{equation}
{H_\textrm{hf}} = {A_\parallel }\left( {{\bf{I}} \cdot {\bf{n}}} \right)\left( {{\bf{J}} \cdot {\bf{n}}} \right) \to \left\langle {\varphi '|{H_\textrm{hf}}|\varphi } \right\rangle  = \frac{{F(F + 1) - I(I + 1) - J(J + 1)}}{{2J(J + 1)}}{A_\parallel }{\Omega ^2}{\delta _{\eta ,\eta '}}
\label{eqn:hyperfine}
\end{equation}
\noindent where $A_\parallel$ is the hyperfine constant and listed in Table \ref{tab:constants}. The matrix elements are independent of the Zeeman sub-levels ($m_\textrm{F}$) and molecular orientation (the sign of $\Omega$). $\eta$ represents all other quantum numbers.

The Stark energy shift comes from the interaction between the molecule-frame dipole moment and DC or quasi-DC electric field, which is defined in the lab frame. Therefore, spherical tensor algebra is required to connect these frames:
\begin{multline}
{H_\textrm{S}}({\bf{E}}) =  - {\bf{d}} \cdot {\bf{E}} = - \sum\limits_p {{{( - 1)}^p}T_p^{(1)}({\bf{d}})T_{ - p}^{(1)}({\bf{E}})} \to \\
\left\langle {\varphi '|{H_\textrm{S}}|\varphi } \right\rangle  =  - d_\textrm{mf} \sum\limits_{p =  - 1}^{ + 1} {{{\left( { - 1} \right)}^p}T_{ - p}^{(1)}({\bf{E}}){{\left( { - 1} \right)}^{F' - {m'_\textrm{F}}}}\left( {\begin{matrix}
{F'} & 1 & F \\
{-{m'_F}} & p & {{m_\textrm{F}}}
\end{matrix}} \right)}\\
\times {\delta _{I,I'}}{\left( { - 1} \right)^{F + J' + 1 + I'}}\sqrt {(2F' + 1)(2F + 1)} \left\{ {
\begin{matrix}
J & F & I' \\
F' & J' & 1
\end{matrix}
} \right\} \\
\times \sum\limits_{q =  - 1}^{ + 1} {{{\left( { - 1} \right)}^{J' - \Omega '}}\left( {
\begin{matrix}
J' & 1 & J \\
-\Omega & q & \Omega
\end{matrix}
} \right)\sqrt {(2J' + 1)(2J + 1)} T_q^{(1)}({\bf{n}})} 
\end{multline}
\noindent where $d_\textrm{mf}$ is the molecule-frame electric dipole moment of $^3\Delta_1$ state of $^{232}$ThF$^+$, which is listed in Table \ref{tab:constants}. $T_p^{(1)}(\bf{V})$ is the spherical components of a rank-1 vector operator. 

In a polar diatomic molecule, such as $^{232}$ThF$^+$, the total angular momentum can have two possible orientations or projections along the inter-nuclear axis, designated as $\pm \Omega$. Due to the coupling of the molecule's electronic and rotational angular momentum, these two orientations do not possess identical energies. Instead, they exhibit a finite energy splitting:
\begin{equation}
{H_\Omega } = \hbar {\omega _\textrm{ef}}{\Omega _\textrm{x}}/2 \to \left\langle {\varphi '|{H_\Omega }|\varphi } \right\rangle  = \frac{{\hbar {\omega _\textrm{ef}}}}{4}J(J + 1){\delta _{\Omega ', - \Omega }}{\delta _{\eta ',\eta }}
\label{eqn:Omega}
\end{equation}
\noindent where $\omega _\textrm{ef}$ is the $\Omega$-doubling constant of $^3\Delta_1$ state of $^{232}$ThF$^+$, which is listed in Table \ref{tab:constants}.

Different from typical Stark spectroscopy with a constant electric field, the experimental scheme in this letter requires a rotating electric field with a frequency from 0 to 100 kHz: 
\begin{equation}
{H_\textrm{rot}} =  - \hbar {{\bf{\omega }}_{{\textrm{rot}}}} \cdot {\bf{F}} \to {\omega _\textrm{rot}}\left\langle {\varphi '|{F_\textrm{x}}|\varphi } \right\rangle
=\frac{\sqrt{2}}{2} (F_+ + F_-)
\label{eqn:rotation}
\end{equation}
\noindent where $\omega_\textrm{rot}$ is the angular frequency of the rotating electric field, whose typical range is listed in Table \ref{tab:constants}, and 
\begin{equation}
{F_ \pm } =  \mp {\left( { - 1} \right)^{F' - m'_\textrm{F}}}
\left( {\begin{matrix}
{F'} & 1 & F \\
{-{m'_\textrm{F}}} & \pm 1 & {{m_\textrm{F}}}
\end{matrix}} \right)
\sqrt{F(F+1)(2F+1)}
\delta_{J,J'} \delta_{\Omega, \Omega'} \delta_{F,F'} \delta{\eta,\eta'}.
\end{equation}

The Zeeman shift comes from two different contributions: (1) electronic and (2) nuclear spin. The Zeeman shift comes from the electronic contribution is:

\begin{multline}
{H_\textrm{Z,e}} =  - {{\bf{\mu }}_{\textrm{e}}} \cdot {\bf{B}} =  - \sum\limits_p {{{( - 1)}^p}T_p^{(1)}({{\bf{\mu }}_{\bf{e}}})T_{ - p}^{(1)}({\bf{B}})} \to \\
\left\langle {\varphi '|{H_\textrm{Z,e}}|\varphi } \right\rangle  =  - {G_\parallel }{\mu _\textrm{B}} \Omega \sum\limits_{p =  - 1}^{ + 1} {{{\left( { - 1} \right)}^p}T_{ - p}^{(1)}({\bf{B}}){{\left( { - 1} \right)}^{F' - {m'_\textrm{F}}}}\left( {\begin{matrix}
{F'} & 1 & F \\
{-{m'_\textrm{F}}} & p & {{m_\textrm{F}}}
\end{matrix}} \right)}\\
\times {\delta _{I,I'}}{\left( { - 1} \right)^{F + J' + 1 + I'}}\sqrt {(2F' + 1)(2F + 1)} \left\{ {
\begin{matrix}
J & F & I' \\
F' & J' & 1
\end{matrix}
} \right\} \\
\times \sum\limits_{q =  - 1}^{ + 1} {{{\left( { - 1} \right)}^{J' - \Omega '}}\left( {
\begin{matrix}
J' & 1 & J \\
-\Omega & q & \Omega
\end{matrix}
} \right)\sqrt {(2J' + 1)(2J + 1)} T_q^{(1)}({\bf{n}})} 
\label{eqn:Zeeman-e}
\end{multline}

\noindent And the Zeeman shift comes from the nuclear spin is:
\begin{multline}
{H_\textrm{Z,N}} =  - {{\bf{\mu }}_{\textrm{N}}} \cdot {\bf{B}} =  - \sum\limits_p {{{( - 1)}^p}T_p^{(1)}({{\bf{\mu }}_{\textrm{N}}})T_{ - p}^{(1)}({\bf{B}})} \to \\
\left\langle {\varphi '|{H_\textrm{Z,N}}|\varphi } \right\rangle  =  - \sum\limits_{p =  - 1}^{ + 1} {{{\left( { - 1} \right)}^p}T_{ - p}^{(1)}({\bf{B}}){{\left( { - 1} \right)}^{F' - {m'_\textrm{F}}}}\left( {\begin{matrix}
{F'} & 1 & F \\
{-{m'_\textrm{F}}} & p & {{m_\textrm{F}}}
\end{matrix}} \right)}\\
\times {\delta _{J,J'}}{\left( { - 1} \right)^{F' + J' + 1 + I}}\sqrt {(2F' + 1)(2F + 1)} \left\{ {
\begin{matrix}
I & F & J' \\
F' & I & 1
\end{matrix}
} \right\} \\
\times \sqrt {I(I + 1)(2I + 1)} g_\textrm{N} \mu_\textrm{N}
\label{eqn:Zeeman-N}
\end{multline}

In the equations (\ref{eqn:Zeeman-e}) and (\ref{eqn:Zeeman-N}), the parameters $G_\parallel$ and $g_\textrm{N}$ signify the strength of the Zeeman interactions. Instead of adopting the theoretical values of these parameters, we use the experimental results from the JILA eEDM group. They have determined an effective magnetic g-factor, $g_F$, for the $^3\Delta_1$, $F=3/2$ states, where the matrices for $H_\textrm{Z,e}$ and $H_\textrm{Z,N}$ coincide. The g-factor for the $^3\Delta_1$, $F=1/2$ states should theoretically differ, albeit within the same order of magnitude. However, no experimental measurements have been conducted for these particular states as of now. Given that the uncertainty of the g-factor for these states shows lower sensitivity to the quantum control scheme discussed in this letter, we use the same $g_\textrm{F}$ for all sub-levels of $^3\Delta_1$. This approximation should not introduce significant discrepancies in our quantum control model.

\begin{table}
\begin{center}
\begin{tabular}{l l l} 
\hline\hline
 Constant & Values & Description  \\
\hline
$B_e/h$ & 7.293(2) GHz & Rotation constant \\ 
${A_\parallel }/h$ & -20.1(1) MHz & Hyperfine constant \\ 
$d_{mf}$ & 3.37(9) Debye & Molecule-frame electric dipole moment \\
$\omega_{ef}/(2\pi)$ & 5.29(5) MHz & $\Omega$-doubling constant \\
$\omega_{rot}/(2\pi)$ & 0-100 kHz & Frequency range of the rotating E-field \\
$g_F$ & 0.0149(3) & $g$-factor of $F=3/2$ state \\
$r_{trap}$ & 3 mm & Trap radius \\
\hline\hline
\end{tabular}
\end{center}
\caption{\label{tab:constants} Constants used in the effective Hamiltonian of ThF$^+$ and spin-polarization calculations.}
\end{table}

\section{Section 5: Results of spin-polarization calculations.}

\begin{figure}[h]
\includegraphics[width=0.9\textwidth]{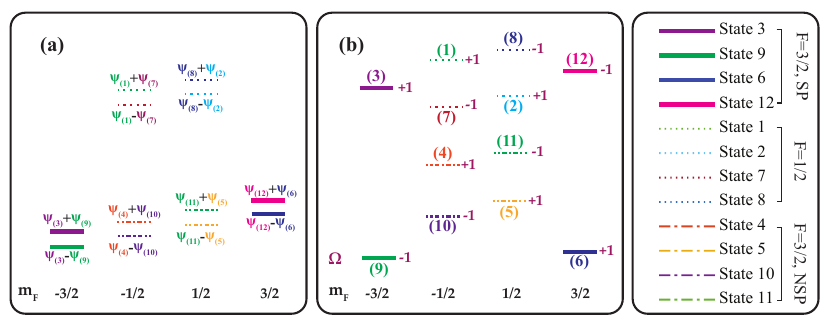}
\caption{\label{fig:state} Depiction of quantum states of $^{232}$ThF$^+$, $^3\Delta_1$, $v=0$, $J=1$. Panel (a) offers a qualitative illustration of ions situated within a static frame without the presence of an electric field. Panel (b) presents a qualitative representation of ions within a rotating frame subjected to a net electric field. The numbering, line styles, and color schemes associated with each state maintain consistency with those used in FIGs \ref{fig:B1} through \ref{fig:B5}. }
\end{figure}

\begin{figure}[H]
\includegraphics[width=0.75\textwidth]{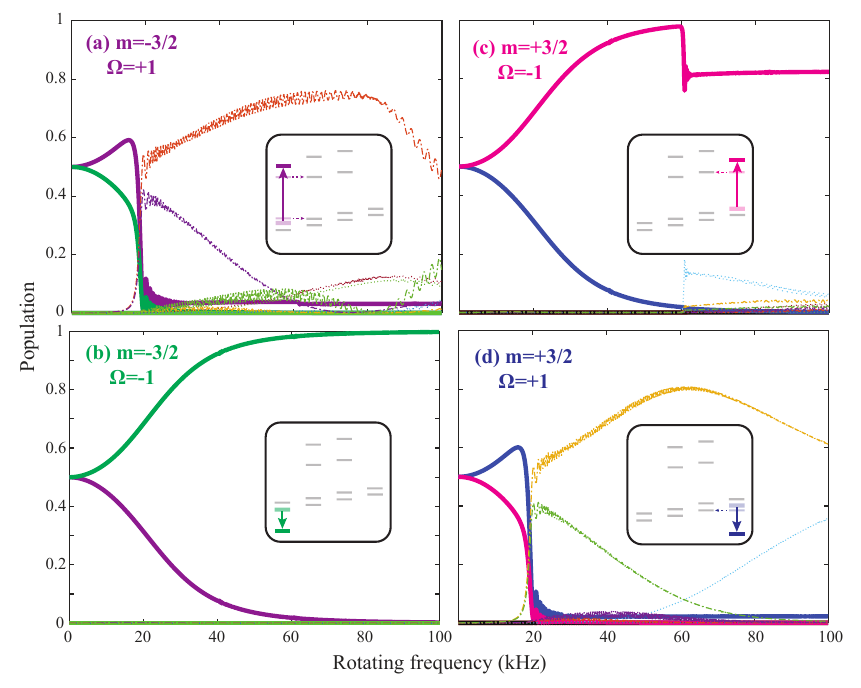}
\centering
\caption{\label{fig:B1} The evolution of quantum states during a frequency ramp-up in the rotating frame over a span of 1 millisecond. Significantly, the population in the $m_\textrm{F}=-3/2$, $\Omega=-1$ state emerges as the sole group that can be preserved throughout this procedure. The dashed arrows featured in the inset diagrams effectively illustrate the pathways of population leakage.}
\end{figure}

We have investigated various quantum control schemes aimed at the preparation of spin-polarized states in the context of transitioning from a static to a rotating frame. FIG \ref{fig:state} shows the energy diagram of $^3\Delta_1$ states in the static frame with zero electric field (panel a) and in the rotating frame with a net 30 V/cm electric field (panel b). In the subsequent computational analysis, we will employ the molecular orientation $\Omega$ as a label for quantum states. It's worth noting that $\Omega$ serves as a good quantum number exclusively under the condition of a sufficiently large electric field. Consequently, in the static frame without an electric field, all states become a superposition of $\Omega=\pm1$.

FIG \ref{fig:B1} graphically depicts the evolution of quantum states, beginning from spin-polarized states where $m_\textrm{F}=\pm3/2$, over the course of a 1 ms frequency ramp-up in the rotating frame. Notably, only the population of the $m_\textrm{F}=-3/2$, $\Omega=-1$ state remains intact. All other spin-polarized states experience either total or partial population losses to alternative states, driven by rotationally induced adiabatic interactions. 

\begin{figure}[H]
\includegraphics[width=0.75\textwidth]{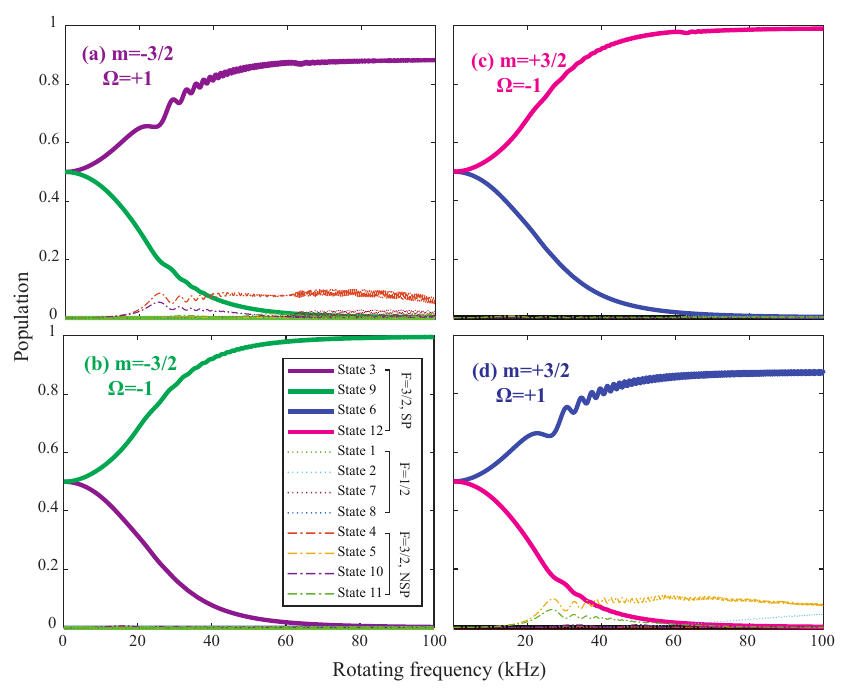}
\centering
\caption{\label{fig:B2} The evolution of quantum states during a frequency ramp-up in the rotating frame over a span of 30 microseconds. }
\end{figure}

FIG \ref{fig:B2} illustrates quantum state evolution from the same initial states as in FIG \ref{fig:B1}, but with a much faster frequency ramp-up rate (from 0 to 100 kHz in 30 $\mu$s). As we accelerate the ramping rate, there's a substantial decrease in the adiabatic population transfer to non-spin-polarizing states. However, achieving such a high acceleration rate proves to be a significant challenge within the confines of a surface ion trap characterized by a relatively shallow trap depth. Additionally, the fidelity of both states, $m_\textrm{F}=-3/2$, $\Omega = +1$ and $m_\textrm{F}=+3/2$, $\Omega=-1$, does not meet our established criteria. 

\begin{figure}[H]
\includegraphics[width=0.75\textwidth]{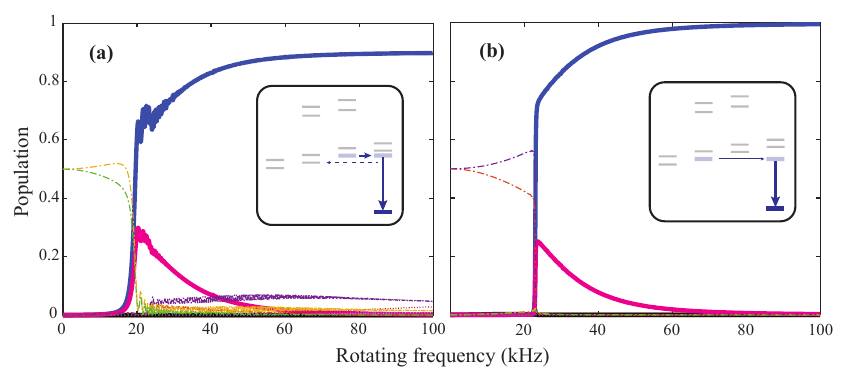}
\centering
\caption{\label{fig:B3} The preparation of the $m_\textrm{F}=3/2$, $\Omega=+1$ state via adiabatic population transfer. In panel (a), the process begins with the $m_\textrm{F}=1/2$, $F=3/2$ state and leverages a first-order interaction over a brief 1 ms rotation ramp-up process. Conversely, panel (b) initiates from the $m_\textrm{F}=-1/2$, $F=3/2$ state and employs a third-order interaction over a considerably longer 75 ms rotation ramp-up period.}
\end{figure}

Rather than preparing the spin-polarized quantum state in the static frame, we propose an alternative approach: preparing the non-spin-polarizing state using the QLS method in the static frame. Subsequently, the population can be adiabatically transitioned to the spin-polarized state during the rotating ramp-up process via rotation-induced coupling. FIG \ref{fig:B3} demonstrates two strategies for preparing the $m_\textrm{F}=3/2$, $\Omega=+1$ state via adiabatic population transfer. In panel (a), we initiate from the $m_\textrm{F}=1/2$, $F=3/2$ state and employ a first-order rotation-induced interaction over a 1 ms rotation ramp-up process. Here, we observe that approximately 90$\%$ of the population is successfully transferred to the target state. However, around 10$\%$ of the population leaks to the $m_\textrm{F}=-1/2$, $F=3/2$ state through a third-order interaction (indicated by a dashed arrow), which includes one Stark coupling and two rotational couplings. Contrastingly, panel (b) commences from the $m_\textrm{F}=-1/2$, $F=3/2$ state and employs a third-order interaction (indicated by a solid arrow) to transfer the population to the target state. This process necessitates a significantly slower ramp-up rate of 75 ms. Remarkably, this method enables the achievement of a 100$\%$ population transfer to the target state. 

\begin{figure}[H]
\includegraphics[width=0.75\textwidth]{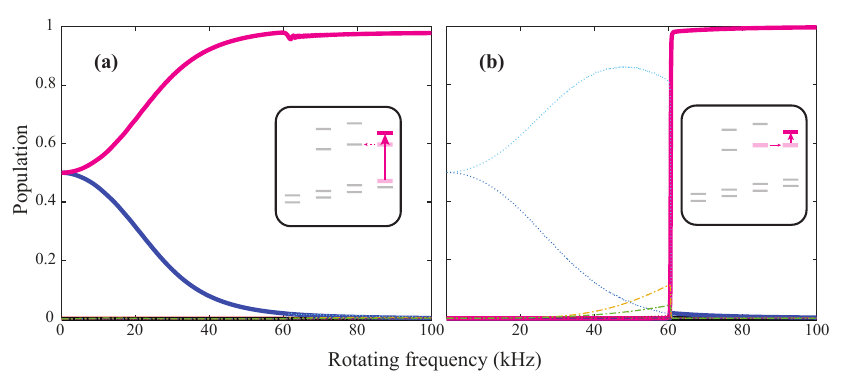}
\centering
\caption{\label{fig:B4} The preparation of the $m_\textrm{F}=3/2$, $\Omega=-1$ state. In panel (a), we commence with the spin-polarized $m_\textrm{F}=+3/2$ state and carry out a swift 0.1 ms rotation ramp-up process to circumvent the adiabatic population transfer to the $F=1/2$ state. Conversely, panel (b) initiates the process from the $F=1/2$ state and employs a comparatively lengthy 75 ms rotation ramp-up process to achieve the adiabatic population transfer.}
\end{figure}

In the case of the $m_\textrm{F}=+3/2$, $\Omega=-1$ state, the Stark interaction causes an upward shift in the state, precluding any chance for degenerate interaction with the $m_\textrm{F}=+1/2$, $F=3/2$ state. However, as the electric field escalates to 12 V/cm at a 60 kHz rotating frequency, population leakage occurs via a relatively weak, yet not insignificant interaction with the $m_\textrm{F}=+1/2$, $F=1/2$ state. A possible solution to mitigate this leakage is to increase the ramp-up rate (to 0.1 ms ramped-up duration), which can result in a transfer efficiency exceeding 95$\%$, as illustrated in FIG \ref{fig:B4}(a). Nonetheless, achieving such a rapid ramp-up rate presents considerable experimental challenges. An alternate and potentially more effective approach begins with the $m_\textrm{F}=+1/2$, $F=1/2$ state and adiabatically transfers the population to the $m_\textrm{F}=+3/2$, $\Omega=-1$ state. As demonstrated in FIG \ref{fig:B4}(b), this method can yield a near 100$\%$ efficiency.

\begin{figure}[H]
\includegraphics[width=0.75\textwidth]{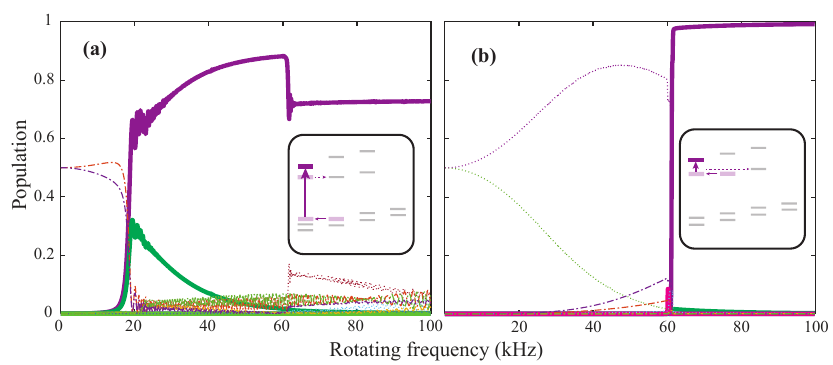}
\centering
\caption{\label{fig:B5} The preparation of the $m_\textrm{F}=-3/2$, $\Omega=+1$ state. Panel (a) begins with the spin-polarized $F=3/2$ state and executes a 1 ms rotation ramp-up process to avert the adiabatic population transfer to the $F=1/2$ state. In contrast, panel (b) initiates from the $F=1/2$ state and employs a more extended 40 ms rotation ramp-up process to facilitate adiabatic population transfer.}
\end{figure}

In the case of the $m_\textrm{F}=-3/2$, $\Omega=+1$ state, we can initiate the process from the $m_\textrm{F}=-1/2$, $F=3/2$ state as depicted in FIG \ref{fig:B5}(a). The population can then be transitioned to the target state via an adiabatic interaction, employing a moderately low ramp-up rate of 1 ms. However, a drawback to this approach is that the population in the $m_\textrm{F}=-3/2$, $\Omega=+1$ state tends to leak to the $F=1/2$ state through an adiabatic interaction at a 60 kHz rotating frequency. An enhanced strategy involves starting with the $m_\textrm{F}=-1/2$, $F=1/2$ state, as illustrated in FIG \ref{fig:B5}(b). This enables the population to be adiabatically transferred to the target state. Despite this improvement, the population still risks leakage to the $m_\textrm{F}=+1/2$, $F=1/2$ state via a high-order interaction (indicated by a dashed arrow). Therefore, in this scenario, it becomes necessary to fine-tune the ramp-up rate for the sake of achieving an efficiency of state preparation that approaches unity.

\section{Supplementary Reference}
\noindent [S1]~Zhou, Y., Shagam, Y., Cairncross, W. B., Ng, K. B., Roussy, T. S., Grogan, T., Boyce, K., Vigil, A., Pettine, M., Zelevinsky, T., Ye, J., and Cornell, E. A. Second-scale coherence
measured at the quantum projection noise limit with hundreds of molecular ions. \textit{Physical Review Letters},
124(5):53201, 2020.

\noindent [S2]~ Shi, M., Herskind, P. F., Drewsen, M. and Chuang, I. L. Microwave quantum logic spectroscopy and control of molecular ions. \textit{New Journal of Physics}, 15, 113019, 2013.

\noindent [S3]~Cairncross, W. B., Gresh, D. N., Grau, M., Cossel, K. C., Roussy, T. S., Ni, Y., Zhou, Y.,
Ye, J., and Cornell, E. A. Precision measurement of the electron’s electric dipole moment using trapped
molecular ions. \textit{Physical Review Letters}, 119(15):153001, 2017.

\noindent [S4]~Ng, K. B., Zhou, Y., Cheng, L., Schlossberger, N., Park, S. Y., Roussy, T. S., Caldwell, L., Shagam, Y., Vigil, A. J., Cornell, E. A.,
and Ye, J. Spectroscopy on the electron-electric-dipole-moment-sensitive states of ThF$^+$. \textit{Physical Review A}, 105(2):022823, 2022.

\end{document}